\documentclass[10pt,twocolumn,letterpaper]{article}

\usepackage{cvpr}
\usepackage{times}
\usepackage{epsfig}
\usepackage{graphicx}
\usepackage{amsmath}
\usepackage{amssymb}
\usepackage{authblk}

\usepackage[pagebackref=true,breaklinks=true,letterpaper=true,colorlinks,bookmarks=false]{hyperref}

\cvprfinalcopy 

\def\cvprPaperID{27} 

\ifcvprfinal\fi
\begin{document}

 \title{SpaceNet 6: Multi-Sensor All Weather Mapping Dataset}
\author[1]{Jacob Shermeyer}
\author[1]{Daniel Hogan}
\author[2]{Jason Brown}
\author[1]{Adam Van Etten}
\author[1]{Nicholas Weir}
\author[3]{Fabio Pacifici}
\author[4]{Ronny Hänsch}
\author[5]{Alexei Bastidas}
\author[2]{Scott Soenen}
\author[3]{Todd Bacastow}
\author[1]{Ryan Lewis}
\affil[1]{In-Q-Tel - CosmiQ Works, [jshermeyer, dhogan, avanetten, nweir, rlewis]@iqt.org}
\affil[2]{Capella Space, [jason.brown, scott]@capellaspace.com}
\affil[3]{Maxar Technologies, Todd.Bacastow@maxar.com}
\affil[4]{German Aerospace Center, ronny.haensch@dlr.de}
\affil[5]{Intel AI Lab, alexei.a.bastidas@intel.com}

\maketitle
\thispagestyle{empty}

\begin{abstract}
   
   Within the remote sensing domain, a diverse set of acquisition modalities exist, each with their own unique strengths and weaknesses. Yet, most of the current literature and open datasets only deal with electro-optical (optical) data for different detection and segmentation tasks at high spatial resolutions. optical data is often the preferred choice for geospatial applications, but requires clear skies and little cloud cover to work well. Conversely, Synthetic Aperture Radar (SAR) sensors have the unique capability to penetrate clouds and collect during all weather, day and night conditions. Consequently, SAR data are particularly valuable in the quest to aid disaster response, when weather and cloud cover can obstruct traditional optical sensors. Despite all of these advantages, there is little open data available to researchers to explore the effectiveness of SAR for such applications, particularly at very-high spatial resolutions, i.e. $<1m$ Ground Sample Distance (GSD).
   
   To address this problem, we present an open Multi-Sensor All Weather Mapping (MSAW) dataset and challenge, which features two collection modalities (both SAR and optical). The dataset and challenge focus on mapping and building footprint extraction using a combination of these data sources. MSAW covers $120 km^2$ over multiple overlapping collects and is annotated with over $48,000$ unique building footprints labels, enabling the creation and evaluation of  mapping algorithms for multi-modal data. We present a baseline and benchmark for building footprint extraction with SAR data and find that state-of-the-art segmentation models pre-trained on optical data, and then trained on SAR ($F_1$ score of 0.21) outperform those trained on SAR data alone ($F_1$ score of 0.135).

\end{abstract}

\begin{figure*}
  \includegraphics[width=\textwidth]{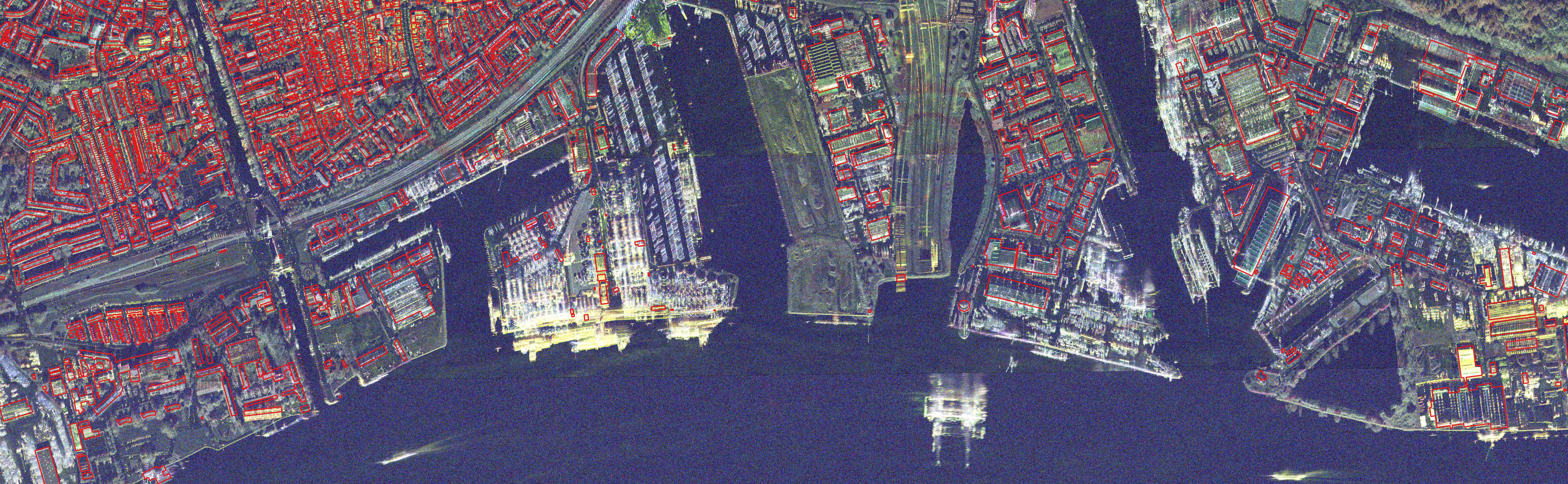}
  \label{header_image}
  \caption{\textbf{Building footprints outlined in red overlaid upon visualized SAR Intensity with three polarizations (HH, VV, HV) displayed through the Red, Green, and Blue color channels.}}
\end{figure*}

\section{Introduction and Related Work}

The advancement of object detection and segmentation techniques in natural scene images has been driven largely by permissively licensed open datasets. For example, significant research has been galvanized by datasets such as ImageNet \cite{imagenet}, MSCOCO \cite{coco} and PASCALVOC \cite{pascalvoc}, among others.  Additionally, multi-modal datasets continue to be developed, with a major focus on 3D challenges, such as PASCAL3D+ \cite{pascal3D}, Berkeley MHAD \cite{ofli2013berkeley}, Falling Things \cite{tremblay2018falling}, or ObjectNet3D \cite{xiang2016objectnet3d}.  Other modalities such as radar remain generally unexplored with very few ground based radar datasets, most of which are focused on autonomous driving such as EuRAD \cite{meyer2019automotive} and NuScenes \cite{caesar2019nuscenes}.  Although these datasets are immensely valuable, the models derived from them do not transition well to the unique context of overhead observation.  Analyzing overhead data typically entails detection or segmentation of small, high-density, visually heterogeneous objects (e.g. cars and buildings) across broad scales, varying geographies, and often with limited resolution - challenges rarely presented by natural scene data. 

Ultimately, few high-resolution overhead datasets exist for mapping and detection of objects in overhead imagery.  The majority of these datasets are specifically focused on leveraging electro-optical (optical) imagery. For example, the permissively licensed SpaceNet \cite{Etten2018SpaceNetAR,Weir_2019_ICCV} corpus presently covers 10 cities, with $~27,000 km^2$ of optical imagery, $>800,000$ building footprints, and $~20,000$ km of road labels.  Less permissively licensed datasets such as xView \cite{Lam:2018}, xBD \cite{xbd}, A Large-scale Dataset for Object DeTection in Aerial Images (DOTA) \cite{dota}, and Functional Map of the World (FMOW) \cite{fmow} are impressively expansive, each addressing different tasks using optical data. However, lacking from each of these datasets are other modalities of data common to remote sensing. One of the most prominent overhead sensor types is synthetic aperture radar (SAR). 

 SAR sensors collect data by actively illuminating the ground with radio waves rather than utilizing the reflected light from the sun as with passive optical sensors.  The sensor transmits a wave, it bounces off of the surface, and then returns back to the sensor (known as backscatter) \cite{moreira2013}.  Consequently, SAR sensors succeed where optical sensors fail: They do not require external illumination and can thus collect at night. Additionally, radar waves pierce clouds, enabling visualization of Earth's surface in all weather conditions. SAR data differs greatly from optical. For example, the intensity of the pixels in a radar image are not indicative of an object's visible color, but rather represent how much radar energy is reflected to the sensor. Reflection strength provides insights on the material properties or physical shape of an object. Depending on the target properties and the imaging geometry, the radar antenna will receive all, some, or none of the radio wave’s energy \cite{moreira2013}. Furthermore, SAR sensors can transmit in up to 4 polarizations  by transmitting in a horizontal or vertical direction and measuring only the horizontally- or vertically-polarized (HH, HV, VH, VV) part of the echo. Each polarization can help distinguish features on the ground by measuring the most prevalent types of scattering for objects of interest \cite{moreira2013}. 
 
\begin{figure}
\vspace{-0mm}
\begin{center}
\includegraphics[width=0.85\linewidth]{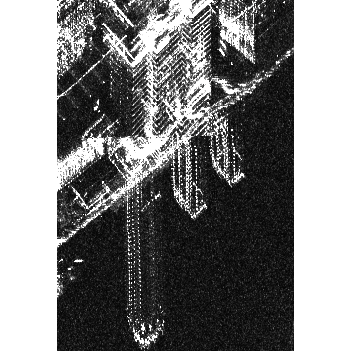}
\end{center}
\caption{\textbf{Examples of layover in urban SAR imagery.} In this detail of Rotterdam, there is land to the north (top) and water to the south (bottom). This image was recorded by aircraft from the south. The three skyscrapers near the riverbank appear to jut into the water because of layover.}
\label{layover_fig}
\vspace{-4mm}
\end{figure}

 SAR imagery presents unique challenges for both computer vision algorithms and human comprehension.  In particular, SAR imagery is considered a non-literal imagery type because it does not look like an optical image which is generally intuitive to humans. These aspects must be understood for accurate image interpretation to be performed. Because SAR is a side-looking, ranging instrument, the backscattered returns will be arranged in the image based on how far the target is from the antenna along the slant plane (radar-image plane). This causes some interesting geometrical distortions in the imagery, such as foreshortening or layover. Tall objects with a slope, such as a mountain, do appear steeper, with a thin bright “edge” appearance at the peak. Layover is an extreme example of foreshortening where the object is so tall that the radar signal reaches the top of an object before it reaches the bottom of it. This causes the returns from the top of the structure to be placed on the image closer to the sensor (near range) and obscure the bottom (Figure \ref{layover_fig}). Such complex geometric issues will present a unique challenge to computer-vision algorithms to comprehend and interpret.
 
  A few SAR-specific datasets exist. Notably, the Moving and Stationary Target Acquisition and Recognition (MSTAR) dataset \cite{diemunsch1998moving} focuses on classifying military vehicles.  The recently released SARptical \cite{sarptical} dataset also focuses on SAR and optical data fusion for foundational mapping purposes.  However, both SARptical and MSTAR are distributed in small tiles, are non-georeferenced, and lack scalability to broader areas. Coarser datasets such as the Sen12MS dataset \cite{sen12ms} provide a valuable global dataset of multi-spectral optical and SAR imagery as well as land-cover labels at $10m$ resolution spanning hundreds of locations.  While such coarser resolution datasets are incredibly useful, to our knowledge no high-resolution ($<=$ 1~m GSD) multi-modal SAR datasets are publicly available with permissive licenses (the 2012 IEEE Geoscience and Remote Sensing Society (GRSS) Data Fusion challenge \cite{berger2013multi} built an excellent dataset, using a combination of high-resolution SAR, optical, and LIDAR over downtown San-Francisco; however these data have a limited license and are no longer publicly available).

\begin{figure*}
  \includegraphics[width=\textwidth]{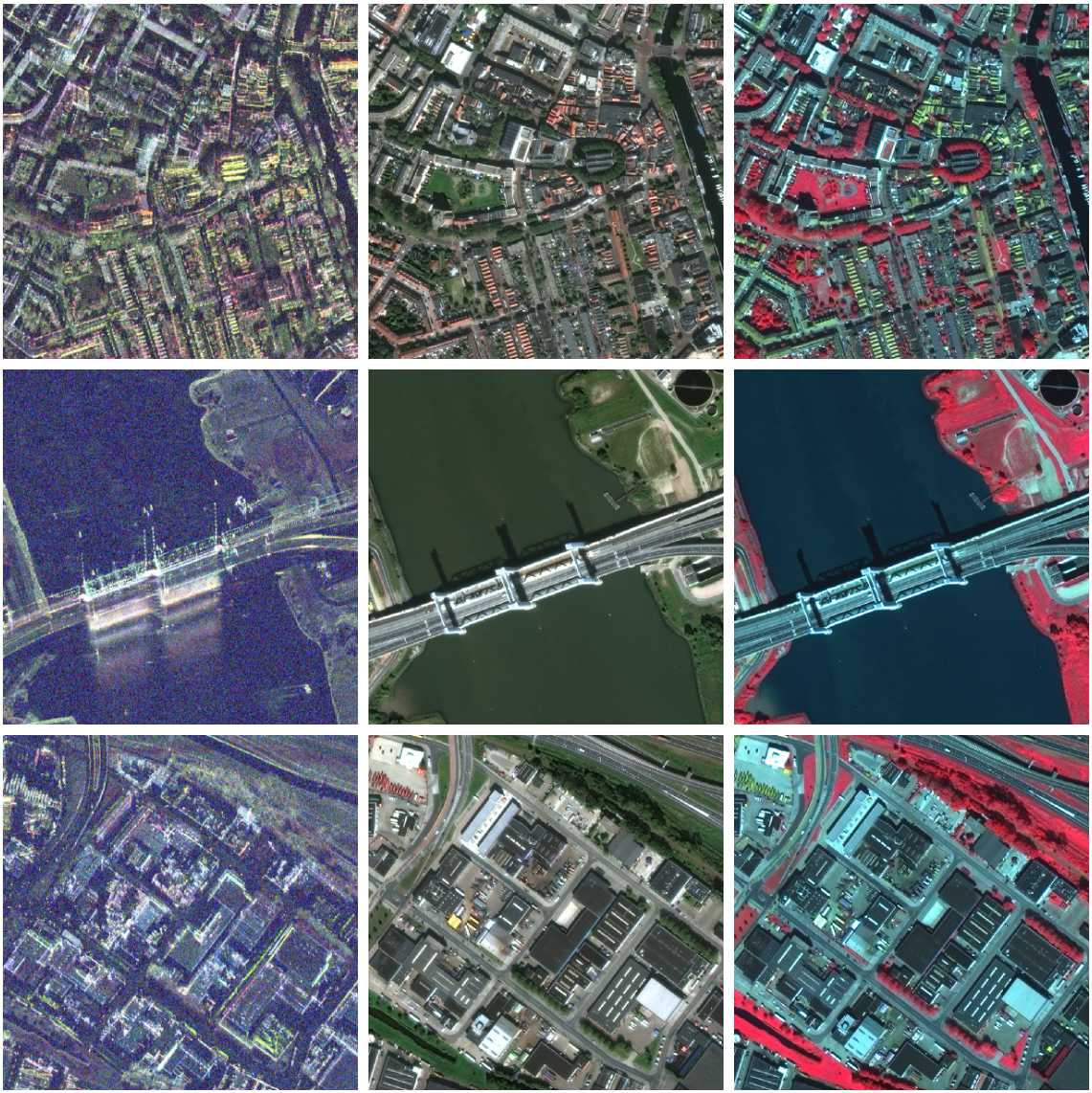}
  \caption{\textbf{Three areas from the SpaceNet 6 MSAW dataset:} Left: SAR Intensity in decibels (HH, VV, VH). Center: Visible Spectrum Imagery (R,G,B). Right: False Color Composite Imagery (NIR, R, G). }
\end{figure*}

  Although SAR has existed since the 1950s \cite{moreira2013} and studies with neural nets date back at least to the 1990s \cite{chen1996}, the first application of deep neural nets to SAR was less than five years ago \cite{morgan2015}.  Progress has been rapid, with accuracy on the MSTAR dataset rising from $92.3\%$ to $99.6\%$ in just three years \cite{morgan2015, furukawa2018}.  The specific problem of building footprint extraction from SAR imagery has been only recently approached with deep-learning \cite{shahzad2019, Xu_buildings}.  Further research is required to investigate the combination of SAR and deep-learning for this task.
 
 To address the limitations detailed above, we introduce the Multi-Sensor All Weather Mapping (MSAW) dataset.  This dataset features a unique combination of half-meter quad-polarized X-band SAR imagery (Figure \ref{header_image}) and half-meter optical imagery over the port of Rotterdam, the Netherlands.  The dataset covers a total area of $120 km^2$ and is labeled with 48,000 unique building footprints, along with associated height information curated from the 3D Basisregistratie Adressen en Gebouwen (3DBAG) dataset \cite{TRIPLE_D_BAG}.  Presently, no other open datasets exist that feature near-concurrent collection of SAR and optical at this scale with sub-meter resolution.  Although limited to a single location, the MSAW dataset represents the first step in creating an openly available very-high resolution repository of SAR data. Moreover, the MSAW dataset joins the existing SpaceNet data corpus, further expanding the geographic diversity and the number of cities to 11. Additionally, we present a deep-learning baseline model for the automated extraction of building footprints using a combination of SAR and optical imagery. Such a baseline is important to demonstrate the performance of state-of-the-art segmentation models for working with SAR data.
 
Alongside the dataset, MSAW also features a public challenge portion, encouraging participants and researchers to produce innovative algorithmic solutions to address challenging foundational mapping problems.  The dataset and challenge results may serve as a baseline and reference benchmark for future research with both overhead SAR and optical imagery. The lessons learned from such a challenge and future experiments with the dataset will provide insights to the broader computer vision community, and enable the design of robust algorithms for a variety of tasks. Finally participants of the challenge and users of the dataset will be invited to participate at the IEEE GRSS EarthVision 2020 workshop (\url{www.grss-ieee.org/earthvision2020/}) to gather, discuss, and help advancing the remote sensing computer vision field. The workshop will take place during the 2020 Conference on Computer Vision and Pattern Recognition (CVPR).

\section{Dataset}

MSAW contains CC-BY-SA 4.0-licensed optical and SAR imagery over the port of Rotterdam, the Netherlands.  Rotterdam is the largest port in Europe, and features thousands of buildings, vehicles, and boats of various sizes, making for an effective test bed for data-fusion experiments between SAR and optical.  MSAW covers an extent of approximately $~120 km^2$. The dataset covers heterogeneous geographies, including high-density urban environments, rural farming areas, suburbs, industrial areas and ports resulting in various building size, density, context and appearance. Additionally, the MSAW dataset is built to mimic real-world scenarios where historical optical data may be available, but concurrent collection of SAR and optical data is not possible.  For example, cloud-cover and adverse weather conditions often complicate remote sensing activities such as disaster response. However, historical high-quality optical data is often available for any area of the earth. As such, the MSAW dataset includes both optical and SAR in the training dataset, but only includes SAR data in the testing dataset. When the dataset is structured in such a fashion the optical data can be used for pre-training or pre-processing, but cannot be used to directly map features. The dataset is available for free download through Amazon Web Services' Open Data Program, with download instructions available at \url{www.spacenet.ai}.

\subsection{Synthetic Aperture Radar}
The SAR data featured in SpaceNet MSAW is provided by Capella Space, in partnership with Metasensing, via an aerial sensor. This sensor mimics the space-borne sensors that will be present on Capella's future constellation of satellites.  The aerial collect captures the same area of Rotterdam multiple times and features 204 individual image strips captured over a three day span: August 4th, 23rd, and 24th 2019. Each strip features four polarizations (HH, HV, VH, and VV) of data in the X-band wavelength. Data are captured from an off-nadir perspective at a relative look angle of 53.4$^{\circ}$ to 56.6$^{\circ}$ from both north- and south-facing directions. These extremely oblique look-angles can present real challenges to traditional computer vision algorithms \cite{Weir_2019_ICCV}. 

The MSAW dataset is processed from Single look complex (SLC) data. By definition SLC data are loosely georeferenced and retain the complex data properties inherent to SAR collections.  We further process these SLC images for each of the four polarizations by co-registering and cropping each to the same extent for each image strip. Next, all polarizations are finely co-registered ($<1$ pixel) to one another using a truncated \emph{sinc} interpolation. The intensity of backscatter (the amount of transmitted radar signal that the imaging surface redirects back to the sensor) is calculated for each polarization. Data are multilooked (noise reduction process) using an average convolution with a $2\times2$ kernel. Any negative intensity value is then set to $0$. Finally, the logarithm of raw intensity is calculated for each polarization and multiplied by 10. Any value that again falls below $0$ is set to $10^{-5}$ and areas containing non-valid imagery are set to $0$. This stretches data to a range between $10^{-5}$ and $92.88$, falling within the 8-bit range of $0 - 255$ to improve usability for challenge participants. These data are then geo-registered and ortho-rectified (correcting imagery for the Earth's complex topography) to the earth's surface using the openly available Shuttle Radar Topography Mission (SRTM) Digital Elevation Model (DEM) and resampled with a lanczos interpolation to a spatial resolution of $0.5 m \times 0.5m$ per pixel.


\subsection{Electro-Optical Imagery}
The optical imagery is provided by the Maxar Worldview-2 satellite. A single, cloud-free image strip was collected on August 31, 2019 at 10:44AM from a look angle of 18.4$^{\circ}$ off-nadir with an overall area of $236km^2$. The collection is composed of three different sets of data with different spatial resolutions:
\begin{itemize}
\item one band panchromatic ($0.5m$)
\item four multi-spectral bands ($2.0m$): blue, green, red, and near-infrared (NIR)
\item four pan-sharpened bands ($0.5m$): blue, green, red, and NIR
\end{itemize}
Pan-sharpening is the process that merges high spatial resolution panchromatic and lower spatial resolution multispectral imagery to create a single high spatial resolution color image. Additionally, each dataset is atmospherically compensated to surface-reflectance values by Maxar's AComp \cite{AComp} and ortho-rectified using the SRTM DEM.  As with the SAR imagery, areas containing non-valid imagery are also set to $0$.

\subsection{Annotations}
We use previously produced high-quality annotations provided openly via the 3DBAG dataset \cite{TRIPLE_D_BAG}.  These labels comprise both building footprints and addresses across all of the Netherlands.  The polygon building footprint labels are produced and curated by the Netherlands Cadastre, Land Registry and Mapping Agency (Kadaster). The dataset is updated regularly as new buildings are registered, built, or demolished.  We use the dataset update from August 28, 2019 containing over 10 million buildings or addresses with $~97\%$ containing valid height information.

The 3D component of the 3DBAG dataset comes from an openly available DEM derived from aerial LiDAR called the Actueel Hoogtebestand Nederland (AHN).  The height information is matched to each building polygon with best-fit RMSE averageing between $25cm-100cm$ based upon a random sample of buildings.  Although the height information will not be used in the challenge, such data can be valuable for future research and analysis on the value of SAR or optical to detect the height of objects from an overhead perspective.

We further refine the dataset by cropping to our area of interest.  Next, we perform manual quality control to add buildings that are missed in the imagery, remove buildings that do not exist in the imagery, and drop certain areas (${<6.5~km^2}$ total) from our training and testing sets where a significant number of buildings are not labeled accurately.  Finally we dissolve individual addresses (i.e. apartments and town-homes) that co-exist within a single building. We then remove buildings smaller than $5 m^2$. The final dataset is comprised of $~$48,000 unique building footprints.

\subsection{Additional Pre-processing}
We tile all data to $450m \times 450m$ ($900 pixels \times 900 pixels$) tiles. We first tile the SAR imagery, and then tile the optical collects to match the corresponding SAR extent and pixel grid.  We then mask the optical data using the extent of each SAR image that contains valid imagery.  Finally we clip our polygon labels to match each tile extent, again removing polygons that do not contain any valid SAR imagery. 

We split the tiled data into three sets: $50\%$ for training, $30\%$ for testing, and $20\%$ for final scoring for the challenge.  The SpaceNet training dataset contains both SAR and optical imagery, however the testing and scoring datasets contain only SAR data. In February 2020 we publicly released SAR, optical, and labels for the training set, SAR only for the public test set, and held back the entire final scoring set.  As mentioned above, we structure the dataset in this way in order to mimic real-world scenarios where historical optical data is available, but concurrent collection with SAR is often not possible due to inconsistent orbits of the sensors, or cloud cover that will render the optical data unusable.

\section{Baseline Building Extraction Experiments}
In conjunction with the MSAW dataset, a baseline algorithm for the MSAW public challenge has been released.  The goal of the challenge is to extract building footprints from SAR imagery, assuming that coextensive optical imagery is available for training data but not for inference.  Releasing a baseline algorithm serves several purposes in the context of the challenge.  First, it demonstrates a complete working solution to the task and illustrates the format for participant submissions.  Second, it gives participants an optional starting point for their own ideas and/or their own code-base.  Third, it serves to set expectations for what is within reach of competitive solutions.

\subsection{Baseline Model}

The baseline algorithm is built around a U-Net \cite{MFing_UNet} neural network with a VGG-11 encoder, an arrangement called a TernausNet\cite{TREX_Net}. The model is trained with an AdamW optimizer \cite{adamw} against a loss function that’s a weighted sum of Dice \cite{dice} and focal loss \cite{focal} which can help with the identification of small objects.  The neural network’s output is a segmentation mask (building or not building), from which individual vector formatted building footprints are extracted. 

It’s possible to train the model to extract building footprints from SAR imagery without using optical data at all. However, model performance can be increased by making use of both modalities. For the baseline, this is done with a transfer learning approach. The model is first trained on optical imagery, then the final weights from that process are used as the initial weights for training on SAR. Since quad-polarization SAR is four-channel, the process is simplified by duplicating a channel in RGB imagery to make it four-channel as well.

To further improve model performance, all SAR imagery is rotated, so that the direction from which the data was collected (and hence the direction of layover) is the same in every case.  Finally, buildings of less than $20 m^2$ (80 total pixels) are ignored during training on SAR data.  These structures, mostly backyard outbuildings, are not readily distinguishable amidst the clutter, and training on their footprints is deleterious to model performance on larger buildings.

\subsection{Metrics}

We used the SpaceNet Metric ($F_1$ score) as defined in Van Etten et al. \cite{Etten2018SpaceNetAR}.  The $F_1$ score uses an intersection-over-union (IOU) threshold to calculate precision $P$ and recall $R$ of model predictions.  We set our IOU threshold at $0.5$ to define a positive detection of a building footprint. 

The SpaceNet Metric is defined as:

\begin{equation}
    F_1 = \frac{2 \times P \times R}{P + R}
\end{equation}

This metric is much more robust versus pixel-wise metrics as it measures the model's ability to delineate building footprints on a per-instance basis, and enables detailed counts of the number and size of buildings present within specific areas.

\subsection{Results and Discussion}

\begin{figure}[b]
\vspace{-0mm}
\begin{center}
\includegraphics[width=0.999\linewidth]{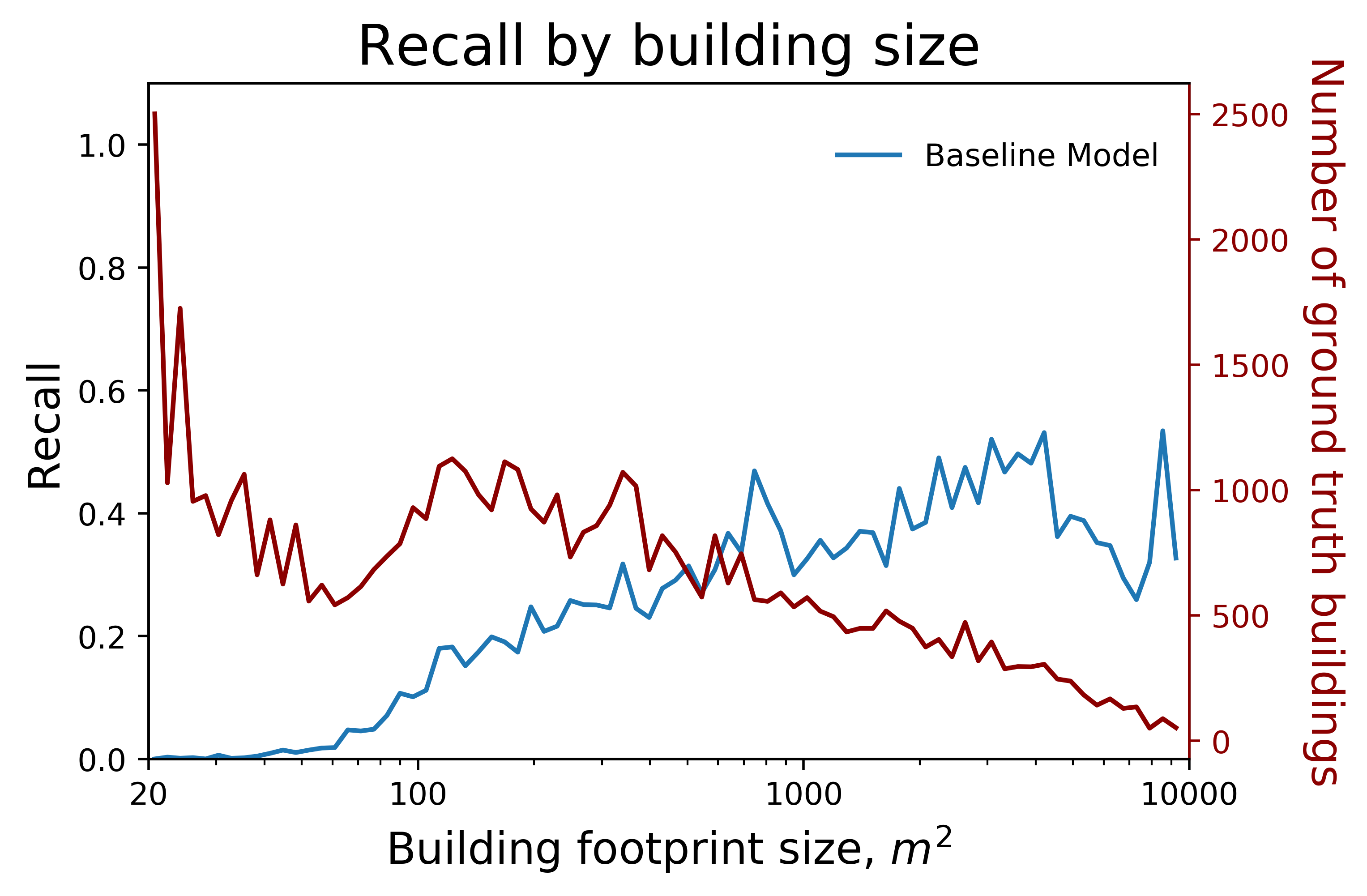}
\end{center}
\caption{\textbf{The effects of building size on model performance (recall).} Recall for the baseline model is plotted in blue with the number of buildings in the dataset by size plotted in red. }
\label{buildings_by_size}
\vspace{-4mm}
\end{figure}

\begin{figure}[b]
\vspace{-0mm}
\begin{center}
\includegraphics[width=0.999\linewidth]{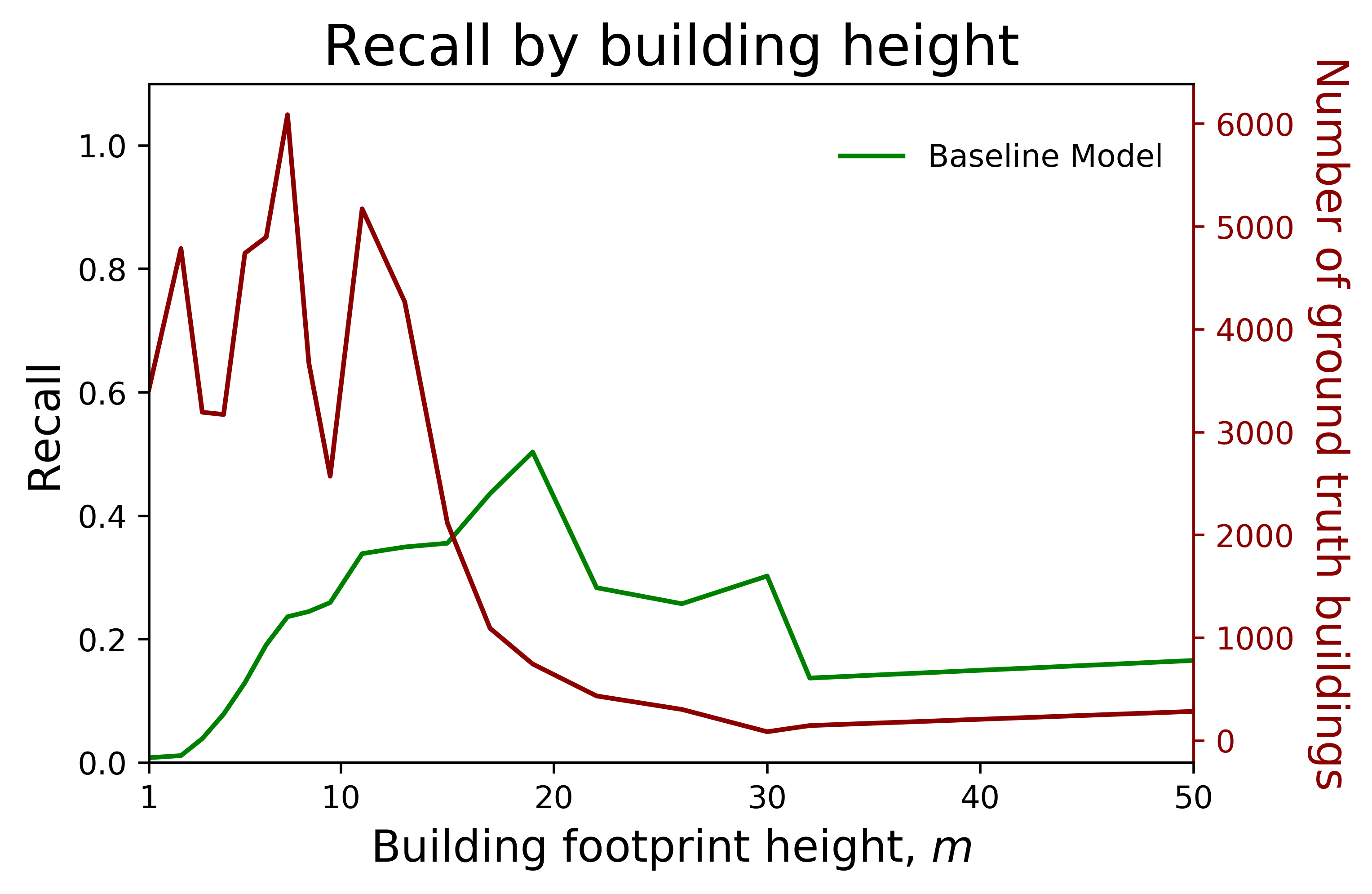}
\end{center}
\caption{\textbf{The effects of building height on model performance (recall).} Recall for the baseline model is plotted in green with the number of buildings in the dataset by height plotted in red. Height is derived from the LiDAR collection associated with the dataset.}
\label{buildings_by_height}
\vspace{-2mm}
\end{figure}

\begin{figure*}
  \includegraphics[width=\textwidth]{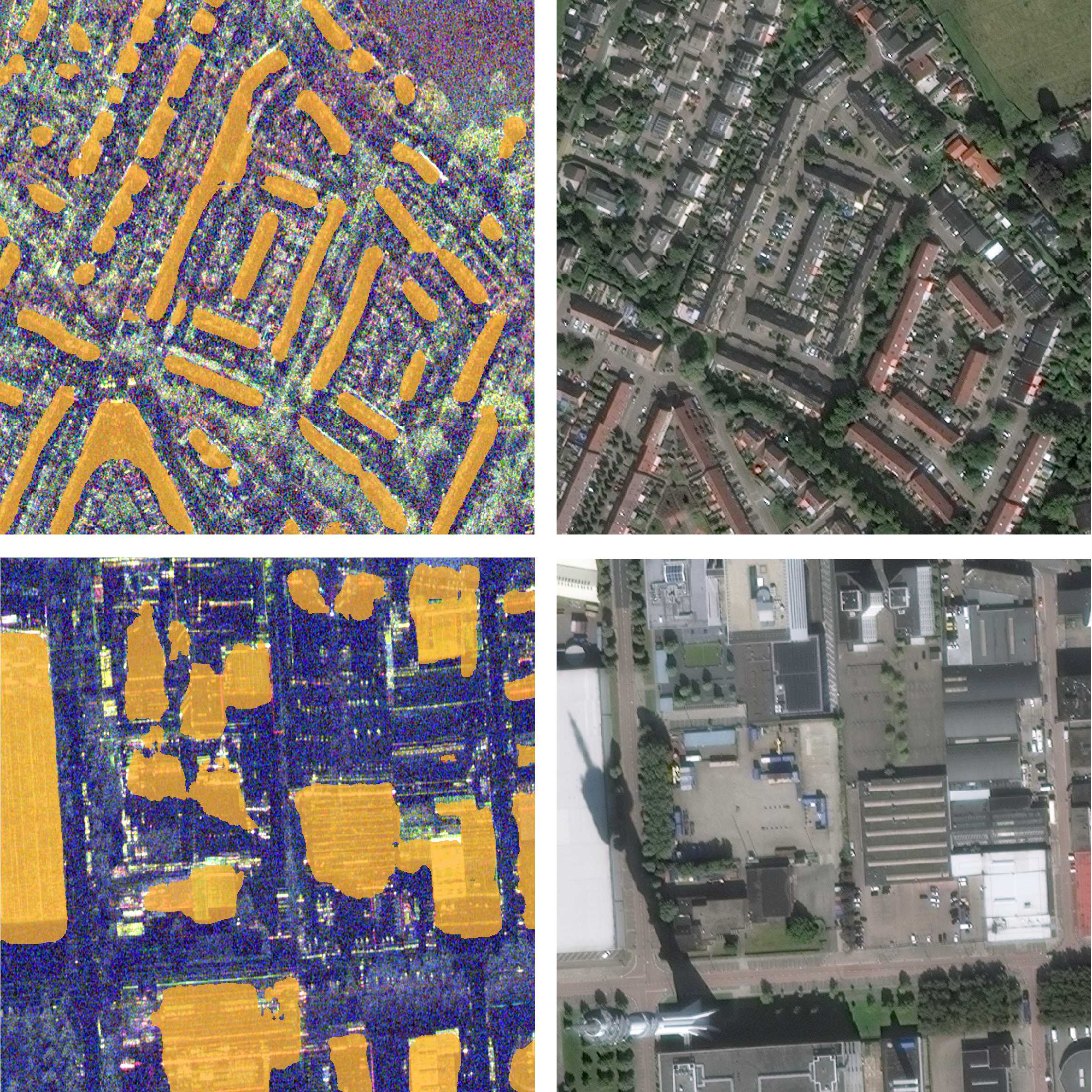}
  \caption{\textbf{Predictions from the MSAW Baseline.} Left: SAR imagery (HH, VV, HV) overlaid with model predictions colorized in orange. Right: Visible spectrum imagery of the same areas. }
  \label{SAR_EO_Preds}
\end{figure*}

Using the metric described above, the baseline algorithm achieves a score of 0.21$\pm$.02 on the MSAW dataset.  For comparison, this is almost identical to the score of a baseline algorithm for a similar building extraction challenge using overhead optical data from a few years ago \cite{Etten2018SpaceNetAR}.  Even recent state-of-the-art models achieve similar performance on optical data that is as far off-nadir as this SAR data \cite{Weir_2019_ICCV}. Figure \ref{SAR_EO_Preds} shows some sample predictions from two testing tiles within the test set.

Score comparison among different versions of the baseline itself show whether different aspects of the baseline design improve performance.  In the absence of transfer learning with optical data, the model’s score drops to 0.135$\pm$.002.  If, in addition, rotational alignment is replaced by random rotations, the score sinks further to 0.12$\pm$.03 (Table \ref{results_table}).  These experimental results show that transfer learning from optical and consistent treatment of SAR viewing direction provide performance benefits.

Ultimately, the baseline algorithm is just one way to approach the problem, and other approaches also merit investigation.  For example, another way to incorporate optical data is a domain adaptation approach where a generative adversarial network is trained to convert SAR images into optical-style images \cite{SAR_TO_EO_1, SAR_TO_EO_2}. SAR to optical image translation is shown to improve land-use classification performance, remove cloud-cover, and boost image quality assessment metrics. Conversely, other studies have found that applying traditional domain adaptation approaches to translate SAR to EO imagery can harm performance for certain tasks \cite{fuentes2019sar}. Ultimately, such approaches are still quite nascent in the SAR domain and further research is required to understand best practices. Alternative approaches, such as existing algorithms for extracting building footprints from SAR \cite{quartulli2004, zhao2013} could be used to generate the neural network’s input, instead of just sending SAR intensity directly into the neural net.  The baseline algorithm is intended as a first step to the broader exploration of solutions in the MSAW public challenge.

We conduct further analysis on the dataset by evaluating model performance based upon building-footprint size (Figure \ref{buildings_by_size}). We find that recall scores achieve a performance of 0.1 at approximately $100 \, \rm{m}^2$.  Any building smaller than this are nearly impossible to detect. Furthermore, we find that performance rises and plateaus at roughly 0.5 recall for much larger buildings, with such buildings becoming increasingly rare in the dataset.  Building height also influences model performance (Figure \ref{buildings_by_height}).  Performance gradually increases as buildings become taller.  We see a peak again around 0.5 recall at a height of approximately 20 meters.  Recall then begins to decline for any buildings larger than 20 meters. The reasons of this could be the results of geometric distortions (layover or foreshortening) which become more extreme as building height increases.  The majority of taller structure ($>30 \, \rm{m}$) have an average area of $1400  \, \rm{m}^2$. Based upon the size analysis, these structures should have an average recall of approximately $0.3$ or greater, however average recall for these taller buildings is actually less than $0.2$.

\begin{table}[]
\label{results_table}
\centering
\begin{tabular}{cc}
\hline
\textbf{Method} & \textbf{$F_1$}   \\ \hline \hline
Optical + SAR        & $0.21 \pm .020$  \\ 
Rotated SAR     & $0.14 \pm .002$ \\
SAR             & $0.12 \pm .030$  \\ \hline
\end{tabular}
\vspace{12pt}
\caption{\textbf{Baseline Performance Results.} We test performance of the baseline and evaluate performance ($F_1$ score) for the semantic segmentation of building footprint in SAR imagery. We report results for a transfer learning approach (optical + SAR), SAR data that has been rotated for a consistent viewing angle (Rotated SAR), and SAR data that has not been rotated to have a consistent viewing angle (SAR).}
\end{table}

\begin{figure}
\vspace{-0mm}
\begin{center}
\includegraphics[width=0.999\linewidth]{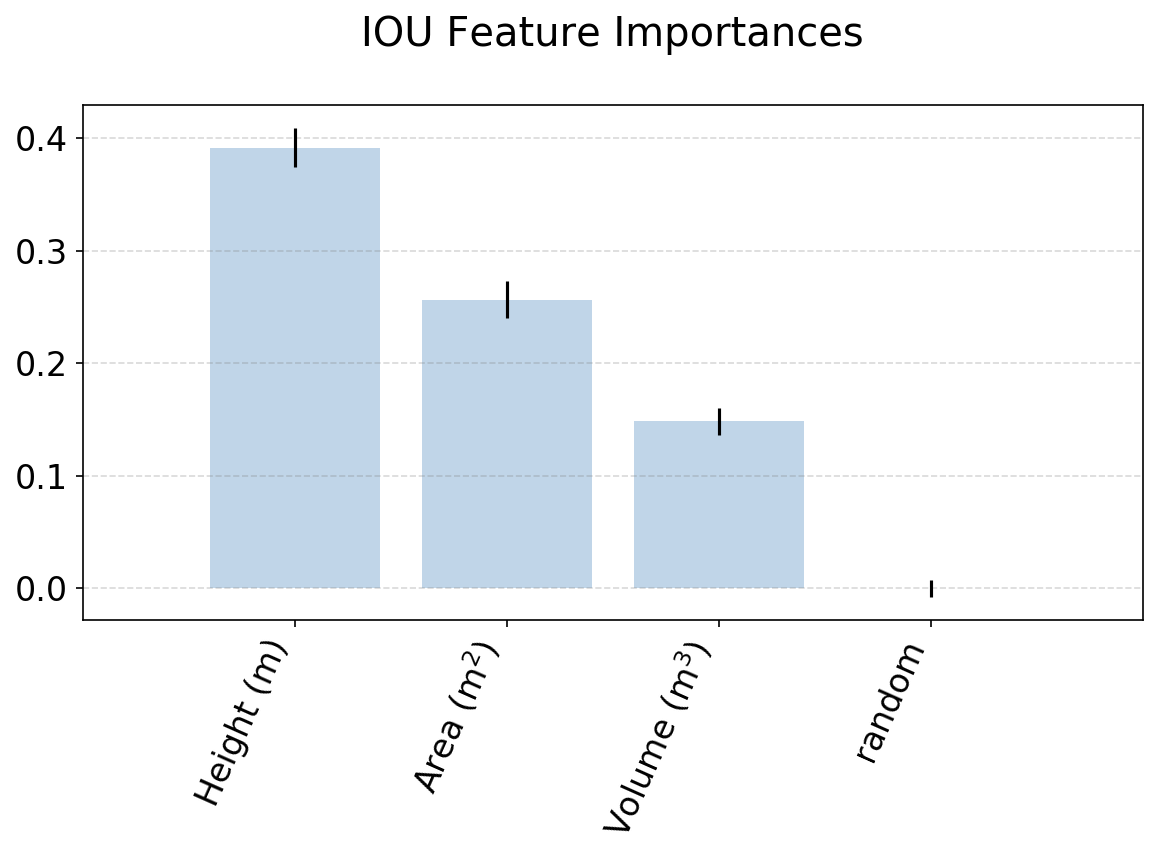}
\end{center}
\vspace{-4pt}
\caption{\textbf{Feature importances}. RF model feature importances for building properties as a predictor of IOU.}
\label{fig:rf_perf}
\vspace{-4mm}
\end{figure}

While building a performant baseline is a difficult and complex task, we can also ask a far simpler question: are building features predictive of the IOU score achieved by the baseline? For this question we follow \cite{sn5_blog} and fit a random forest model to the baseline IOU score using the available building properties (height, area, and volume) as input variables.  This random forest model provides a good fit to the provided data with $\rm{R}^2 = 0.91$.
As expected, a random variable shows no predictive power.  While the Spearman rank correlation coefficients between IOU and height, area, volume are not particularly high 
(+0.16, +0.12, +0.14, respectively) we nevertheless note relatively high predictive feature importance for building height, with a value of 0.39 (out of 1.0), see Figure \ref{fig:rf_perf}.  Even though the baseline model does not explicitly use building height, this feature nevertheless is highly predictive (and positively correlated) with IOU score. This is likely due to the correlation between building height and area (Spearman coefficient of 0.59), with small buildings being much more difficult than larger ones. For very tall buildings the layover greatly complicates building footprint extraction, so height is also predictive of low scores for tall buildings.




\section{Conclusion}
We introduced the Multi-Sensor All Weather Mapping (MSAW) dataset, baseline, and challenge focused on building footprint extraction from an overhead perspective. This novel multi-modal dataset contains both SAR and optical imagery as well as $48,000$ attendant building footprint labels, each tagged with height estimates from a Light Detection and Ranging (LiDAR) sensor. In this paper, we describe the dataset, explain the evaluation metrics, and present a state-of-the-art baseline and quantitative benchmarks for building footprint extraction with SAR and optical data. We find that state-of-the-art segmentation models trained with multiple modalities outperform those trained with only a single type of data. Our experiments indicate that pre-training on optical data and using a transfer learning approach can provide a $55\%$ increase in performance over training on SAR data alone. Regardless of this improvement, the relatively low overall $F_1$ score of $0.21$ showcases the value that future research could provide for extraction of various features in high-resolution SAR data.

Our aim in producing the MSAW dataset is to enable future research around the analysis of multi-modal data and provide techniques that could be applicable in the broader computer vision field. In particular, we hope that MSAW will enable new data-fusion techniques, the evaluation of the detection of small objects, and the testing of domain adaptation algorithms across unique modalities of data. Furthermore, we hope that the further analysis of SAR data will broaden its usefulness in usability, particularly in disaster response scenarios where cloud-cover can often obstruct tradition optical sensors.

{\small
\bibliographystyle{ieee_fullname}
\bibliography{egbib}
}


\end{document}